\documentstyle[aps,prl]{revtex}
\begin{document}
\twocolumn[       %typeset the title and abstract in one column
\title{Higgs Lagrangian from Gauge Theories}
\author{Noriaki Kitazawa\cite{e-mail-kitazawa}
             and
             Francesco Sannino\cite{e-mail-sannino}}
\address{Department of Phyiscs, Yale University,
                New Haven, CT 06520, USA}
\date{\today}
\maketitle
\widetext
\vskip-1.5in
\rightline{\vbox{
\hbox{YCTP-1-98}
\hbox{hep-th/9802017}
}}
\vspace{1in}
\begin{abstract}
\begin{center}
\begin{minipage}{5.5in}
We explore a novel way of deriving 
 the effective Higgs Lagrangian from strongly interacting
 vector-like gauge theories. We consider the 
$N=1$ supersymmetric extension of gauge theories and interpret
 the auxiliary field
 associated with the low energy effective ``meson'' superfield 
as the Higgs field. 
By introducing an explicit supersymmetry breaking term 
and computing the one-loop effective action at the effective theory 
level 
we show that
 the kinetic term for the Higgs field is generated,
 while the negative mass squared term is already present 
 at the tree level.
We further propose a scenario
 by which the complete Higgs potential can be generated 
 and the fermion in the low energy effective theory acquires a mass.
\end{minipage}
\end{center}
\end{abstract}


] % end twocolumn format

\narrowtext

Spontaneous symmetry breaking
 as described by the Higgs Lagrangian (linear $\sigma$-model) 
 has always been the least appealing ingredient of the standard model.
It is a wide spread hope that
 some new and more fundamental gauge dynamics could explain it.  
However it is in general very hard to show
 how the Higgs Lagrangian actually appears
 as a low energy effective theory. 
This problem would indeed require
 to solve the full strong coupling dynamics for gauge theories.
In this letter
 we explore a new method which shows
 how the effective Higgs Lagrangian can emerge from gauge theories.  

To overcame the problem of solving the non-perturbative dynamics,
 we consider the $N=1$ supersymmetric extension
 of vector-like gauge theories.
For these theories
 Seiberg deduced a number of ``exact results''\cite{Seiberg,review}. 
In particular, it was argued that,
 in some cases, the anomalies are sufficient to constrain
 the low energy effective superpotential
 in terms of the lowest-lying composite ``mesonic'' operators
 $M=A + \sqrt{2} \theta \psi + \theta^2 F$ and 
\begin{equation}
 M \sim Q\tilde{Q} \ ,
\label{smeson}
\end{equation}
 where $Q = \phi_{Q}+ \sqrt{2} \theta \psi_{Q} + \theta^2 F_{Q}$ and
 $\tilde{Q} = \phi_{\tilde{Q}}+ \sqrt{2} \theta \psi_{\tilde{Q}}
 + \theta^2 F_{\tilde{Q}}$
 are the vector-like ``quark'' superfields,
 and we have suppressed the flavor indices for simplicity
 while the color sum is understood. 
At the component level, Eq.~(\ref{smeson}) provides
 $A \sim \phi_{Q} \tilde{\phi}_{Q}$,
 $\psi \sim \psi_{Q} \phi_{\tilde{Q}} + \phi_{Q} \psi_{\tilde{Q}}$
 and  
 $F \sim \phi_{Q} F_{\tilde{Q}} + \phi_{\tilde{Q}} F_{Q}
       - \psi_{Q} \psi_{\tilde{Q}}$. 
We can see that
 $A$ and $\psi$ couple to the operators which contain 
at least one squark,
 while only the auxiliary field $F$ is 
associated with the quark pair 
 $\psi_{Q} \psi_{\tilde{Q}}$.  
In the supersymmetric regime 
 the field $F$ must be eliminated via its equation of motion.

Once supersymmetry is broken by introducing a suark mass 
term we expect the operator $F$ to be dominated by the 
$\psi_{Q} \psi_{\tilde{Q}}$ 
operator. 
Under such physical observation
 it seems reasonable to identify the effective auxiliary field $F$ 
with the Higgs field.  
We are now going to show  how an effective Higgs Lagrangian, 
describing the spontaneous breaking of chiral symmetry, 
can actually emerge once supersymmetry is explicitly broken. 

In this letter we concentrate on describing the ideas and final results.
A more complete discussion which also describes the techniques 
involved will be presented in Ref.\cite{paper}.

We consider here as our laboratory
 the asymptotically free supersymmetric gauge theory
 with $N_c=2$ and $N_f=3$.
The global quantum symmetry group is
 $SU\left(2N_f\right) \otimes U_R(1)$,
 and the low energy effective field $M^{ij}$
 belongs to the $15_{\rm A}$ representation of $SU(6)$,
 where $i,j = 1,2, \cdots, 6$ are flavor indices.
The effective low energy superpotential
 which at tree level saturates
 the anomalous Ward-Takahashi identities\cite{Seiberg} is
\begin{equation}
 W=-\frac{1}{\Lambda^3}{\cal P}{\it f} M \, , 
\label{superpotential}
\end{equation} 
 where $\Lambda$ is the dynamical scale associated with 
 the underlying supersymmetric gauge theory. 
In order to define the low energy effective theory 
 we adopt a canonical K\"ahler potential of the form 
\begin{equation}
 K=\frac{1}{\left(\alpha \Lambda\right)^2} M^{\dagger}_{ij} M^{ji} \, ,
\label{naive}
\end{equation}
 where $\alpha$ is an unknown dimensionless parameter of order unity.
Strictly speaking
 the K\"ahler potential should respect scale anomaly
 and hence we might consider a scale invariant K\"ahler potential
 of the type $\sqrt{M^{\dagger}M}$. 
However for simplicity we will not consider such term.

By replacing $M/\left(\alpha \Lambda\right)\rightarrow M$
 the Lagrangian can be cast in a canonical form 
\begin{equation} 
 {\cal L} = \int \! d^4\theta \,M^{\dagger}_{ij} M^{ji} 
 -{\alpha^3}\int \!d^2\theta \,{\cal P}{\it f} M +{\rm h.c.} \, .
\label{clagrangian}
\end{equation}  
At the component level
 the tree level potential from Eq.~(\ref{clagrangian}) is
\begin{equation}
 V\left( F,A \right)
 = - F^{\dagger}_{ij}F^{ji}
   + \frac{\alpha^3}{16} 
      \epsilon_{ijklmn}F^{ij}A^{kl}A^{mn} + {\rm h.c.} \, .
\label{mass}
\end{equation}
The auxiliary field has a negative squared mass
 in the potential of Eq.~(\ref{mass}).
This is consistent with the observation that 
 the auxiliary field in the effective Lagrangian 
 can be the Higgs field associated with the fermion pair condensate
 in the underlying theory.

We now introduce an explicit supersymmetry breaking term which 
in the underlying theory induces a mass term for the squark field:
\begin{equation}
 {\cal L}_{\rm soft}
  = \int d^4 \theta \, X Q^{\dag \alpha}_i Q_\alpha^i \, ,
\label{softmass}
\end{equation}
 where $X = m \theta^2 + m^{\dag} {\bar \theta}^2$
 is a spurion vector superfield
 and we will assume $m$ real. 
$Q_\alpha^i$ is the quark chiral superfield
 with color index $\alpha = 1,2$ and flavor index $i = 1,2, \cdots,6$.
The induced squark mass term preserves chiral symmetry. It is also 
interesting to note that, once the supersymmetry breaking 
term in Eq.~(\ref{softmass})  is  considered,  
the origin of the moduli space, where gauge 
symmetry is preserved,  becomes the unique vacuum 
of the theory. 

In the low energy effective theory we include 
the effects of soft suspersymmetry breaking 
by using the spurion method\cite{soft,soft2}.
Namely we consider adding  the following term 
\begin{equation}
 {\cal L}_{b} = \beta \int \! d^4\theta \, X M^{\dagger}_{ij} M^{ji} \, ,
\label{breaking}
\end{equation}
 to the Lagrangian in Eq.~(\ref{clagrangian}) as first term in a 
supersymmetry breaking expansion ($m < \Lambda$).
Furthermore, in the following, 
 we will partially include non-perturbative supersymmetry 
breaking effects 
by performing the loop calculations 
within the Euclidean momentum cutoff scheme. The latter 
is known to explicitly break supersymmetry.

Now we show
 how an ordinary kinetic term for the auxiliary field is generated
 by evaluating quantum corrections in the effective theory 
framework.
It is clear that
 the two point function for the low energy effective field
 must be responsible for the kinetic term of the auxiliary field. 
The actual one-loop computation yields
\begin{eqnarray}
 \Gamma_2 = \frac{3}{2\pi^2} \left(\frac{\alpha}{2}\right)^6 
 \int \! d^4x \, d^4\theta
 &\biggl[&
  {\rm ln} \left(\frac{\Lambda^2}{m^2}\right) M^{\dagger}_{ij} M^{ji} 
 \nonumber\\
 &+&
 \frac{1}{48 m^2} {\rm \bar{D}}^2 M^{\dagger}_{ij} {\rm D}^2 M^{ji}
 \biggr] \, ,
\label{2PI}
\end{eqnarray}
 where we used the Euclidean momentum cutoff 
regularization scheme and the energy scales $\Lambda$ and $m$
 are respectively identified as the ultraviolet and infrared cutoff.
We remind the reader that
 since supersymmetry is explicitly broken
 we do not need to use a supersymmetry-preserving regularization scheme.
The appearance of controgradient fields (i.e. ${\rm {D}}^2 M$)
 in the second term of Eq.~(\ref{2PI})
 is due to the momentum expansion.
The second term is the source of the kinetic term for the auxiliary field.
\begin{equation}
 {\cal L}_{kin}
  = - \frac{1}{2\pi^2} \left(\frac{\alpha}{2}\right)^6 \frac{1}{m^2} \,
      \partial_m F^{\dagger}_{ij} \partial^{m}F^{ji} \, . 
\label{kinetic}
\end{equation}
(In this letter we adopt the notation of Ref.\cite{Wess-Bagger} )
The first term in Eq.~(\ref{2PI}) provides a correction to the tree 
level negative mass squared term
 for the auxiliary field.
Amusingly it does not have opposite sign
 with respect to the tree level term in Eq.~(\ref{mass}).  
Both terms in Eq.~({\ref{2PI}}) are singular for $m\rightarrow 0$.
 This behavior is the result of our choice of the regularization scheme 
and can be interpreted as a new source of non-perturbative 
supersymmetry breaking 
corrections. 

The Lagrangian for the canonically normalized field 
\begin{equation}
 H = \frac{1}{\sqrt{2}\pi m} \left(\frac{\alpha}{2}\right)^3 \, F \ ,
\end{equation}
 resulted by computing the one-loop two point effective action is   
\begin{equation}
 {\cal L}_{2} = -\partial_m H^{\dagger}_{ij} \partial^m H^{ji} 
                + \mu^2 H^{\dagger}_{ij} H^{ji} \, ,
\label{l2higgs}
\end{equation}
 where 
\begin{equation}
 \mu ^2 = m^2
  \left[
   2\pi^2 \left(\frac{2}{\alpha}\right)^6
   + 3\ln \! \left(\frac{\Lambda^2}{m^2}\right)
  \right] \, .
\end{equation} 
Because of the negative mass  term in Eq.~(\ref{l2higgs}),
 we expect the underlying $SU(6)$ chiral symmetry
 to break spontaneously to $Sp(6)$. In this scenario  
't~Hooft's anomaly matching conditions for chiral anomalies\cite{'tHooft}
 can be saturated by Nambu-Goldstone bosons. 
Before supersymmetry breaking  the 
low energy massless composite fermions  were saturating 
the matching conditions. 
In the following
 we will show that  explicit supersymmetry breaking together with 
spontaneous chiral symmetry breaking will consistently induce 
 a mass term for the composite fermions in the low energy effective theory. 

The one-loop contribution to the effective action
 obtained by computing the four point function is
\begin{eqnarray}
 \Gamma_4
  &=& \frac{1}{2^6 \pi^2} \left(\frac{\alpha^3}{2^3}\right)^4 \frac{1}{m^4}
      \, \int \! d^4x d^4\theta \,
 \nonumber\\
  && \biggl[
   M^{\dagger}_{ij} M^{ji}   \bar{\rm D}^2 M^{\dagger}_{lk} 
         {\rm D}^2 M^{kl}
    + M^{\dagger}_{ij} M^{kl}  \bar{\rm D}^2 M^{\dagger}_{lk} 
{\rm D}^2 M^{ji}
 \nonumber\\
  &~&
    + 8M^{\dagger}_{ji} M^{il} \bar{\rm D}^2 M^{\dagger}_{lk} 
         {\rm D}^2 M^{kj} 
     \biggr]
 \, .
\label{4PI}
\end{eqnarray}
Here we present only the terms relevant to the Higgs Lagrangian.
The resultant contribution in terms of the $H$ field is 
\begin{equation} 
 {\cal L}_{4}
 = \frac{\pi^2}{2^3}
   \left[
     \left(H^{\dagger}_{ij} H^{ji}\right)^2
    + 4 H^{\dagger}_{ij} H^{jl} H^{\dagger}_{lm}H^{mi}
   \right] \, .
\label{l4}
\end{equation}

There are two problems associated with this result.
The first is that 
 there is no dependence on the breaking mass $m$.
Therefore,
 for large values of the parameter $m$, the contribution to the potential is
 suppressed relatively to the effective mass term which scales as $m^2$.   
The second and apparently more serious problem is the fact that  
 the terms presented in Eq.~(\ref{l4}) are positive definite
 and cannot stabilize the potential.  
 We  now suggest a simple way to solve these two problems 
simultaneously.

We notice that
 the result in Eq.~(l4) depends on the choice of the K\"ahler potential.
We might imagine that
 a more complicated K\"ahler potential is needed
 to actually describe the low energy physics.

Here we see that by adding a supersymmetry preserving 
higher dimensional operator to 
the K\"ahler potential we might have the 
correct sign and a better scaling
 with respect to the supersymmetry breaking parameter
 for the $H^4$ term. 
We expect such operator to be dynamically generated 
by the underlying theory and to assume the following form 
at the effective theory level: 
\begin{equation} 
 \frac{\gamma}{\Lambda^2}
  \int \! d^4x d^4\theta \,
  {\rm Tr} \left( M^{\dagger} M M^{\dagger} M \right)
\label{4new}
\end{equation}
 where $M$ is the canonically normalized low energy effective field
 and $\gamma$ is a dimensionless parameter.
The one-loop diagram which contributes 
to the four point function,
 with one vertex generated by Eq.~(\ref{4new}),
  yields
\begin{equation}
 {\cal L}_4^\gamma
  \propto \gamma \frac{m^2}{\Lambda^2}
  {\rm Tr} \left( H^{\dag} H \right)^2 \, .
\end{equation}
Unfortunately we cannot predict the sign of $\gamma$
 and hence we cannot deduce the sign of the full contribution.
If we assume the correct sign for stabilizing the Higgs potential 
the vacuum expectation value for the Higgs field 
is independent of $m$, i.e.
\begin{equation}
 \langle H \rangle \sim \Lambda \, .
\end{equation}
This result is obtained since the 
$H^4$ coupling and mass term
 follow the same scaling with respect to the $m$. 
It is also interesting to observe  that due 
to the present  $(m/\Lambda)^2$ scaling for the $H^4$ coupling 
the latter increases with the supersymmetry 
breaking scale $m$.   

The composite fermion in the low energy effective theory, $\psi$,
 is expected to acquire a mass 
since it contains at least one heavy squark
 and the 't Hooft anomaly matching conditions
 are now saturated via Nambu-Goldstone modes. 
Indeed by an actual one-loop calculation of the four point function,
 we can show that the following terms are generated: 
\begin{equation}
 \frac{\beta \alpha^6}{m}
  H^{\dagger}_{ij} H^{\dagger}_{kl} \psi^{ij} \psi^{kl} \ ,
 \qquad
 \frac{\beta \alpha^6}{m}
  H^{\dagger}_{ik} H^{\dagger}_{jl} \psi^{ij} \psi^{kl} \ .
\label{fmass} 
\end{equation} 

This result shows that when 
chiral symmetry is spontaneously broken
 by $\langle H \rangle \ne 0$,
 the composite fermion acquires a mass.
This feature is similar to the one
 obtained in previous analysis for $N_f<N_c$\cite{soft}
 where the auxiliary field in the effective theory was integrated out.
However we believe that the auxiliary field should 
be kept at 
energies below the scale $m$. 
We notice that
 the radiatively generated mass term in Eq.~(\ref{fmass})
 decreases as the breaking parameter increases.
This dependence can be cured
 by considering the effects of the new K\"ahler term  in 
Eq.~(\ref{4new}). 
In this case the mass dependence is linear in $m$.

We remark that
 the present model for generating the Higgs Lagrangian also illustrates
 how a non-holomorphic low energy effective potential 
might be generated for strong interacting gauge theories. 
(Here by holomorphic
   we mean the potential to be of the form $\chi(\phi) +{\rm h.c.}$,
   where $\chi$ is any function of the generic complex field $\phi$.) 
An holomorphic part of the QCD potential,
 which encodes the anomaly structure of the theory,
 with a suitable procedure for a complete decoupling of supersymmetry
 was derived and studied in Ref.~\cite{toy,anomaly}
 for supersymmetric QCD-like theories with $N_f<N_c$ and $N_c >2$.    

We conclude
 by recapitulating the essential ideas and results presented
 in this letter. 
We suggested a novel way of deriving the Higgs Lagrangian
 from strongly interacting vector-like gauge theories. 
To partially solve the non-perturbative dynamics we considered 
the $N=1$ supersymmetric extension of gauge theories.
It was recognized
 that the auxiliary field for the low energy effective superfield
 can be associated with the quark pair condensate
 which is the order parameter for chiral symmetry breaking.
A supersymmetry breaking mass term for the squark field was introduced to 
recover the original theory (plus a fermion in the adjoint 
representation of the gauge group).

By using as our laboratory the low energy effective Lagrangian 
for the supersymmetric gauge theory with 
$N_c=2$ and $N_f=3$ we uncovered the following features.
We generated at one-loop the kinetic term for 
the auxiliary field (Higgs field) in the effective theory framework,
 while the negative squared mass term was already present 
at tree level.
A stabilizing Higgs potential cannot be generated at one-loop 
if  the naive K\"ahler potential in Eq.~(\ref{naive}) is used. 
However it was shown that a non-trivial K\"ahler potential 
(see Eq.~(\ref{4new})) might solve this problem. 
The massless composite fermion field which saturates 't~Hooft 
anomaly matching conditions in the supersymmetric theory 
acquires a mass (at one-loop level) 
when supersymmetry is broken and if chiral symmetry 
is spontaneously broken by a non-zero vacuum 
expectation value of $H$.
Some non-perturbative supersymmetry breaking effects were 
partially included by performing the loop calculations within 
the Euclidean momentum cutoff regularization scheme, which also 
breaks supersymmetry.

 It would, certainly, be very interesting to consider
 the $N=2$, rather than  $N=1$, supersymmetric extension
 of gauge theories, since the effective 
K\"ahler potential at low energies 
is claimed to be exactly known.

\section*{Acknowledgments}

The  work of N.K. and F.S.  has been partially supported by the US DOE under contract 
DE-FG-02-92ER-40704. The work of N.K. has  
also been supported in part by the Grant-in-Aid for Scientific Research 
from the Ministry of Education, Science, and Culture of Japan
on Priority Areas (Physics of CP violation) and from Inter-Univ. Coop. Research  
under contract \#09045036.

\end{document}